\documentclass[12pt,a4paper]{article}

\usepackage{ifthen} 
\newboolean{pdflatex}
\setboolean{pdflatex}{true} 

\newboolean{articletitles}
\setboolean{articletitles}{true} 

\newboolean{uprightparticles}
\setboolean{uprightparticles}{false} 

\newcommand{\dd}{\ensuremath{{\rm d}}}

\def\BuPiPiPi     {\mbox{\decay{\Bpm}{\pipm\pim\pip}}\xspace}
\def\BuKPiPi      {\mbox{\decay{\Bpm}{\Kpm\pim\pip}}\xspace}
\def\Bpipmmumu{\mbox{{\ensuremath{\Bpm\to\pipm\mup\mun}}}\xspace}
\def\Bpipmumu{{\mbox{\ensuremath{\Bp\to\pip\mup\mun}}}\xspace}
\def\Bpimmumu{{\mbox{\ensuremath{\Bm\to\pim\mup\mun}}}\xspace}
\def\BKpmmumu{{\mbox{\ensuremath{\Bpm\to\Kpm\mup\mun}}}\xspace}

\def\bdmumu{{\mbox{\ensuremath{\bquark\to\dquark\mup\mun}}}\xspace}
\def\bsmumu{{\mbox{\ensuremath{\bquark\to\squark\mup\mun}}}\xspace}
\def\diffBF{{\ensuremath{\left<\frac{\dd\BF}{\dd\qsq}\right>}}\xspace}

\def\BFratio{{\ensuremath{R_{\pion/\kaon}}}\xspace}
\def\BToJPsiK{{\ensuremath{\Bpm\to\jpsi\Kpm}}\xspace}
\def\ACPRaw{{\ensuremath{\mathcal{A}^{C\!P}_\mathrm{raw}}}\xspace}

\def\AProd{{\ensuremath{\mathcal{A}_\mathrm{prod}}}\xspace}
\def\ADet{{\ensuremath{\mathcal{A}_\mathrm{det}}}\xspace}
\def\qsqmin{{\ensuremath{q^2_\mathrm{min}}}\xspace}
\def\qsqmax{{\ensuremath{q^2_\mathrm{max}}}\xspace}

\def\paperauthors{LHCb collaboration} 
\def\paperasciititle{Differential measurements of the branching fraction and CP asymmetry of the decay B->pimumu} 
\def\papertitle{Differential measurements of the branching fraction and \CP asymmetry of the decay \Bpipmmumu} 
\def\paperkeywords{{High Energy Physics}, {LHCb}} 
\def\papercopyright{\the\year\ CERN for the benefit of the LHCb collaboration} 
\def\paperlicence{CC BY 4.0 licence}
\def\paperlicenceurl{https://creativecommons.org/licenses/by/4.0/}

\newif\ifEnableSectionTOCLinks
\EnableSectionTOCLinksfalse 

\usepackage[top=1in, bottom=1.25in, left=1in, right=1in]{geometry}

\usepackage{microtype}
\usepackage{lineno}  
\usepackage{xspace} 
\usepackage{caption} 

\usepackage{graphicx}  
\usepackage{color}
\usepackage{colortbl}
\graphicspath{{./figs/}} 

\usepackage{amsmath} 
\usepackage{amssymb}
\usepackage{amsfonts}
\usepackage{upgreek} 

\newcommand*\patchAmsMathEnvironmentForLineno[1]{%
\expandafter\let\csname old#1\expandafter\endcsname\csname #1\endcsname
\expandafter\let\csname oldend#1\expandafter\endcsname\csname
end#1\endcsname
 \renewenvironment{#1}%
   {\linenomath\csname old#1\endcsname}%
   {\csname oldend#1\endcsname\endlinenomath}%
}
\newcommand*\patchBothAmsMathEnvironmentsForLineno[1]{%
  \patchAmsMathEnvironmentForLineno{#1}%
  \patchAmsMathEnvironmentForLineno{#1*}%
}
\AtBeginDocument{%
\patchBothAmsMathEnvironmentsForLineno{equation}%
\patchBothAmsMathEnvironmentsForLineno{align}%
\patchBothAmsMathEnvironmentsForLineno{flalign}%
\patchBothAmsMathEnvironmentsForLineno{alignat}%
\patchBothAmsMathEnvironmentsForLineno{gather}%
\patchBothAmsMathEnvironmentsForLineno{multline}%
\patchBothAmsMathEnvironmentsForLineno{eqnarray}%
}

\usepackage[pdftex,
            pdfauthor={\paperauthors},
            pdftitle={\paperasciititle},
            pdfkeywords={\paperkeywords}]{hyperref}
\usepackage{hyperxmp}
\hypersetup{
    pdfcopyright={Copyright (C) \papercopyright},
    pdflicenseurl={\paperlicenceurl}
}

\usepackage[colorinlistoftodos,textsize=scriptsize]{todonotes}

\usepackage[bottom,flushmargin,hang,multiple]{footmisc}

\usepackage[all]{hypcap} 

\usepackage{xspace} 
\usepackage{upgreek}

\def\lhcb   {\mbox{LHCb}\xspace}

\def\MagUp {\mbox{\em Mag\kern -0.05em Up}\xspace}

\ifthenelse{\boolean{uprightparticles}}%
{

 \def\Pmu         {\ensuremath{\upmu}\xspace}

 \def\Ppi         {\ensuremath{\uppi}\xspace}

 \def\Ppsi        {\ensuremath{\uppsi}\xspace}

 \def\PDelta      {\ensuremath{\Delta}\xspace}                 
 \def\PXi         {\ensuremath{\Xi}\xspace}                 
 \def\PLambda     {\ensuremath{\Lambda}\xspace}                 
 \def\PSigma      {\ensuremath{\Sigma}\xspace}                 
 \def\POmega      {\ensuremath{\Omega}\xspace}                 
 \def\PUpsilon    {\ensuremath{\Upsilon}\xspace}
 \let\oldPi\Pi
 \def\PPi         {\ensuremath{\oldPi}\xspace}

 \def\PB      {\ensuremath{\mathrm{B}}\xspace}                 
 \def\PD      {\ensuremath{\mathrm{D}}\xspace}                 
 \def\PJ      {\ensuremath{\mathrm{J}}\xspace}                 
 \def\PK      {\ensuremath{\mathrm{K}}\xspace}                 
 \def\Pb      {\ensuremath{\mathrm{b}}\xspace}                 
 \def\Pc      {\ensuremath{\mathrm{c}}\xspace}                 
 \def\Pd      {\ensuremath{\mathrm{d}}\xspace}

 \def\Ph      {\ensuremath{\mathrm{h}}\xspace}                 
 \def\Ps      {\ensuremath{\mathrm{s}}\xspace}                 
 \def\Pt      {\ensuremath{\mathrm{t}}\xspace}                 

 \def\thebaroffset{0.0em}
}
{

 \def\Pmu         {\ensuremath{\mu}\xspace}

 \def\Ppi         {\ensuremath{\pi}\xspace}

 \def\Ppsi        {\ensuremath{\psi}\xspace}                 
                  
 \mathchardef\PDelta="7101
 \mathchardef\PXi="7104
 \mathchardef\PLambda="7103
 \mathchardef\PSigma="7106
 \mathchardef\POmega="710A
 \mathchardef\PUpsilon="7107
 \mathchardef\PPi="7105
 \def\PB      {\ensuremath{B}\xspace}                 
 \def\PD      {\ensuremath{D}\xspace}                 
 \def\PJ      {\ensuremath{J}\xspace}                 
 \def\PK      {\ensuremath{K}\xspace}                 
 \def\Pb      {\ensuremath{b}\xspace}                 
 \def\Pc      {\ensuremath{c}\xspace}                 
 \def\Pd      {\ensuremath{d}\xspace}

 \def\Ph      {\ensuremath{h}\xspace}                 
 \def\Ps      {\ensuremath{s}\xspace}                 
 \def\Pt      {\ensuremath{t}\xspace}                 

 \def\thebaroffset{0.18em}
}
\newcommand{\offsetoverline}[2][\thebaroffset]{\kern #1\overline{\kern -#1 #2}}%

\makeatletter
\ifcase \@ptsize \relax
  \newcommand{\miniscule}{\@setfontsize\miniscule{4}{5}}
\or
  \newcommand{\miniscule}{\@setfontsize\miniscule{5}{6}}
\or
  \newcommand{\miniscule}{\@setfontsize\miniscule{5}{6}}
\fi
\makeatother

\DeclareRobustCommand{\optbar}[1]{\shortstack{{\miniscule (\rule[.5ex]{1.25em}{.18mm})}
  \\ [-.7ex] $#1$}}

\def\mup        {{\ensuremath{\Pmu^+}}\xspace}
\def\mun        {{\ensuremath{\Pmu^-}}\xspace} 

\def\dquark    {{\ensuremath{\Pd}}\xspace}

\def\squark    {{\ensuremath{\Ps}}\xspace}

\def\cquark    {{\ensuremath{\Pc}}\xspace}

\def\bquark    {{\ensuremath{\Pb}}\xspace}

\def\tquark    {{\ensuremath{\Pt}}\xspace}

\def\pion   {{\ensuremath{\Ppi}}\xspace}

\def\pip    {{\ensuremath{\pion^+}}\xspace}
\def\pim    {{\ensuremath{\pion^-}}\xspace}
\def\pipm   {{\ensuremath{\pion^\pm}}\xspace}

\def\kaon    {{\ensuremath{\PK}}\xspace}

\def\KorKbar {\kern \thebaroffset\optbar{\kern -\thebaroffset \PK}{}\xspace}

\def\Kp      {{\ensuremath{\kaon^+}}\xspace}
\def\Km      {{\ensuremath{\kaon^-}}\xspace}
\def\Kpm     {{\ensuremath{\kaon^\pm}}\xspace}

\def\KS      {{\ensuremath{\kaon^0_{\mathrm{S}}}}\xspace}

\def\Kstar   {{\ensuremath{\kaon^*}}\xspace}

\def\D       {{\ensuremath{\PD}}\xspace}

\def\DorDbar {\kern \thebaroffset\optbar{\kern -\thebaroffset \PD}\xspace}
\def\Dz      {{\ensuremath{\D^0}}\xspace}

\def\Dp      {{\ensuremath{\D^+}}\xspace}
\def\Dm      {{\ensuremath{\D^-}}\xspace}

\def\DpDm    {\ensuremath{\Dp {\kern -0.16em \Dm}}\xspace}

\def\B       {{\ensuremath{\PB}}\xspace}

\def\BorBbar {\kern \thebaroffset\optbar{\kern -\thebaroffset \PB}\xspace}
\def\Bz      {{\ensuremath{\B^0}}\xspace}

\def\Bd      {{\ensuremath{\B^0}}\xspace}

\def\BdorBdbar {\kern \thebaroffset\optbar{\kern -\thebaroffset \Bd}\xspace}
\def\Bu      {{\ensuremath{\B^+}}\xspace}
\def\Bub     {{\ensuremath{\B^-}}\xspace}
\def\Bp      {{\ensuremath{\Bu}}\xspace}
\def\Bm      {{\ensuremath{\Bub}}\xspace}
\def\Bpm     {{\ensuremath{\B^\pm}}\xspace}

\def\Bs      {{\ensuremath{\B^0_\squark}}\xspace}

\def\BsorBsbar {\kern \thebaroffset\optbar{\kern -\thebaroffset \Bs}\xspace}

\def\Bcpm    {{\ensuremath{\B_\cquark^\pm}}\xspace}

\def\jpsi     {{\ensuremath{{\PJ\mskip -3mu/\mskip -2mu\Ppsi}}}\xspace}
\def\psitwos  {{\ensuremath{\Ppsi{(2S)}}}\xspace}

\def\Y#1S{\ensuremath{\PUpsilon{(#1S)}}\xspace}

\def\LorLbar     {\kern \thebaroffset\optbar{\kern -\thebaroffset \PLambda}\xspace}

\def\BF         {{\ensuremath{\mathcal{B}}}\xspace}

\newcommand{\decay}[2]{\ensuremath{\mathinner{#1\!\to #2}}\xspace}

\def\to                 {\ensuremath{\rightarrow}\xspace}

\def\qsq       {{\ensuremath{q^2}}\xspace}

\def\CP                {{\ensuremath{C\!P}}\xspace}

\def\Vtd  {{\ensuremath{V_{\tquark\dquark}^{\phantom{\ast}}}}\xspace}

\def\Vts  {{\ensuremath{V_{\tquark\squark}^{\phantom{\ast}}}}\xspace}

\newcommand{\ACP}{{\ensuremath{{\mathcal{A}}^{\CP}}}\xspace}

\def\bsll     {\decay{\bquark}{\squark \ell^+ \ell^-}}

\def\AT#1     {\ensuremath{A_{\mathrm{T}}^{#1}}\xspace}           

\def\C#1      {\ensuremath{\mathcal{C}_{#1}}\xspace}                       
\def\Cp#1     {\ensuremath{\mathcal{C}_{#1}^{'}}\xspace}                    
\def\Ceff#1   {\ensuremath{\mathcal{C}_{#1}^{\mathrm{(eff)}}}\xspace}        
\def\Cpeff#1  {\ensuremath{\mathcal{C}_{#1}^{'\mathrm{(eff)}}}\xspace}       
\def\Ope#1    {\ensuremath{\mathcal{O}_{#1}}\xspace}                       
\def\Opep#1   {\ensuremath{\mathcal{O}_{#1}^{'}}\xspace}                    

\newcommand{\nospaceunit}[1]{\ensuremath{\text{#1}}}       
\newcommand{\aunit}[1]{\ensuremath{\text{\,#1}}}       

\newcommand{\tev}{\aunit{Te\kern -0.1em V}\xspace}
\newcommand{\gev}{\aunit{Ge\kern -0.1em V}\xspace}
\newcommand{\mev}{\aunit{Me\kern -0.1em V}\xspace}
\newcommand{\kev}{\aunit{ke\kern -0.1em V}\xspace}
\newcommand{\ev}{\aunit{e\kern -0.1em V}\xspace}
 
\newcommand{\mevc}{\ensuremath{\aunit{Me\kern -0.1em V\!/}c}\xspace}
\newcommand{\gevc}{\ensuremath{\aunit{Ge\kern -0.1em V\!/}c}\xspace}
\newcommand{\mevcc}{\ensuremath{\aunit{Me\kern -0.1em V\!/}c^2}\xspace}
\newcommand{\gevcc}{\ensuremath{\aunit{Ge\kern -0.1em V\!/}c^2}\xspace}
\newcommand{\gevgevcccc}{\ensuremath{\gev^2\!/c^4}\xspace} 

\def\mum  {\ensuremath{\,\upmu\nospaceunit{m}}\xspace}

\def\fb   {\ensuremath{\aunit{fb}}\xspace}
\def\invfb   {\ensuremath{\fb^{-1}}\xspace}

\def\gsim{{~\raise.15em\hbox{$>$}\kern-.85em
          \lower.35em\hbox{$\sim$}~}\xspace}
\def\lsim{{~\raise.15em\hbox{$<$}\kern-.85em
          \lower.35em\hbox{$\sim$}~}\xspace}

\def\pt         {\ensuremath{p_{\mathrm{T}}}\xspace}

\def\ptot       {\ensuremath{p}\xspace}

\def\evtgen     {\mbox{\textsc{EvtGen}}\xspace}

\def\geant      {\mbox{\textsc{Geant4}}\xspace}

\def\photos     {\mbox{\textsc{Photos}}\xspace}

\def\pythia     {\mbox{\textsc{Pythia}}\xspace}

\def\roofit     {\mbox{\textsc{RooFit}}\xspace}

\def\tell1  {TELL1\xspace}
\def\ukl1   {UKL1\xspace}

\newcommand{\lhcborcid}[1]{\href{https://orcid.org/#1}{\hspace*{0.1em}\raisebox{-0.45ex}{\includegraphics[width=1em]{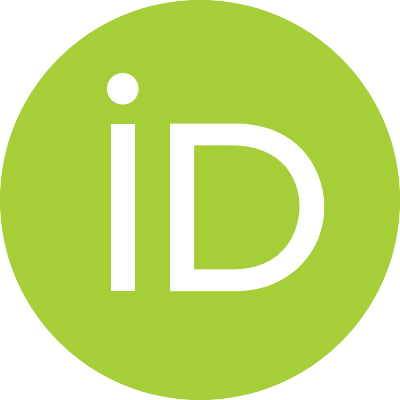}}}}

\hypersetup{
  colorlinks   = true, 
  urlcolor     = blue, 
  linkcolor    = blue, 
  citecolor    = red   
}

\ifEnableSectionTOCLinks
    \usepackage[explicit]{titlesec} 
    
    \let\oldcontentsline\contentsline
    \renewcommand

    \titleformat{\section}{\normalfont\Large\bf}{\hyperlink{tocsection.\thesection}{{\thesection} \parbox[t]{\dimexpr\textwidth-1pc}{#1}}}{1pc}{}

    \titleformat{\subsection}{\normalfont\bf}{\hyperlink{tocsubsection.\thesubsection}{{\thesubsection} \parbox[t]{\dimexpr\textwidth-1pc}{#1}}}{1pc}{}

    \titleformat{name=\section,numberless}[display]{}{}{0pt}{\normalfont\Huge\bfseries #1}
\fi

\usepackage{cite} 
\usepackage{LHCb/mciteplus}

\usepackage{longtable} 
\usepackage{multirow}
\usepackage{graphicx}
\usepackage{subcaption}
\usepackage{booktabs}

\begin{document}

\renewcommand{\thefootnote}{\fnsymbol{footnote}}
\setcounter{footnote}{1}


\begin{titlepage}
\pagenumbering{roman}

\vspace*{-1.5cm}
\centerline{\large EUROPEAN ORGANIZATION FOR NUCLEAR RESEARCH (CERN)}
\vspace*{1.5cm}
\noindent
\begin{tabular*}{\linewidth}{lc@{\extracolsep{\fill}}r@{\extracolsep{0pt}}}
\ifthenelse{\boolean{pdflatex}}
{\vspace*{-1.5cm}\mbox{\!\!\!\includegraphics[width=.14\textwidth]{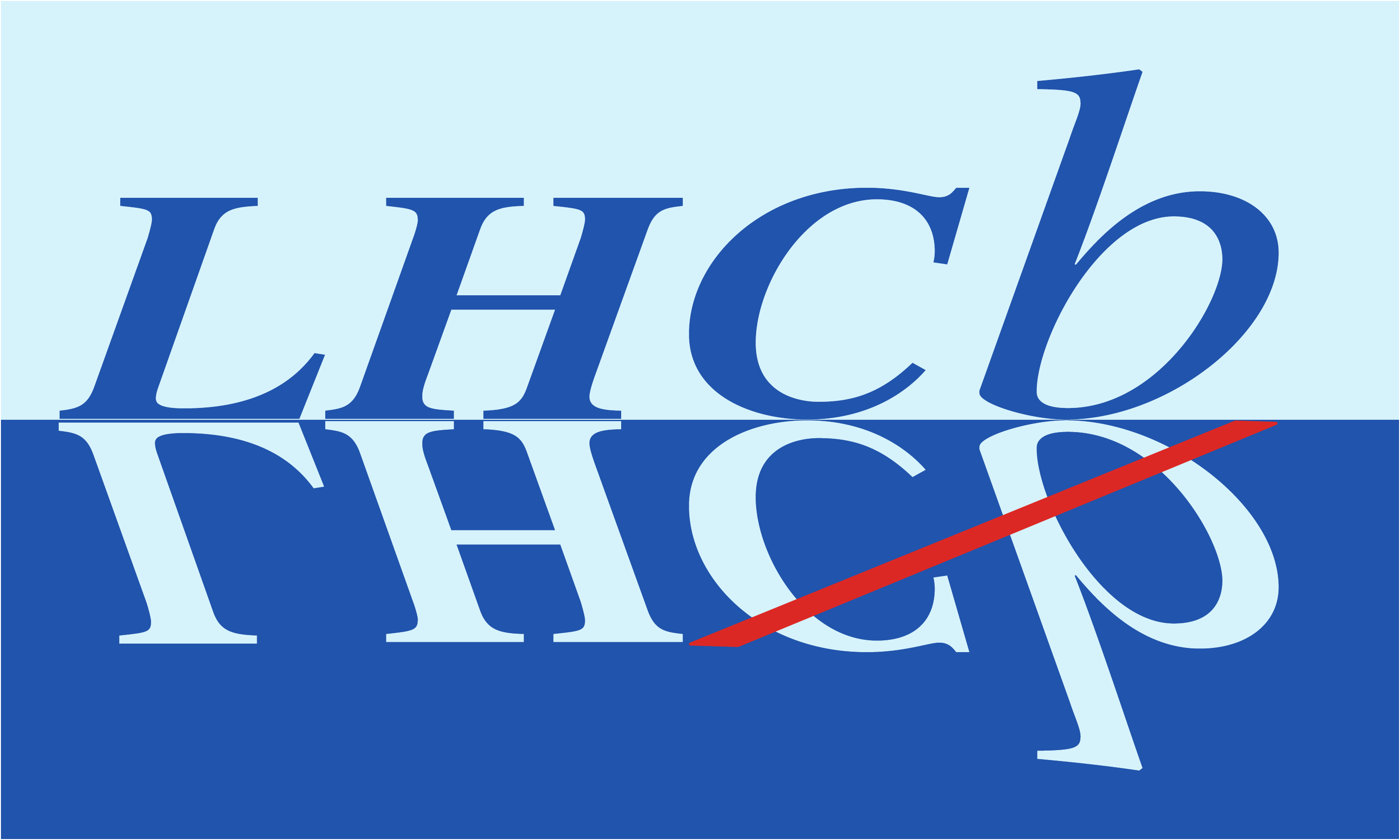}} & &}%
{\vspace*{-1.2cm}\mbox{\!\!\!\includegraphics[width=.12\textwidth]{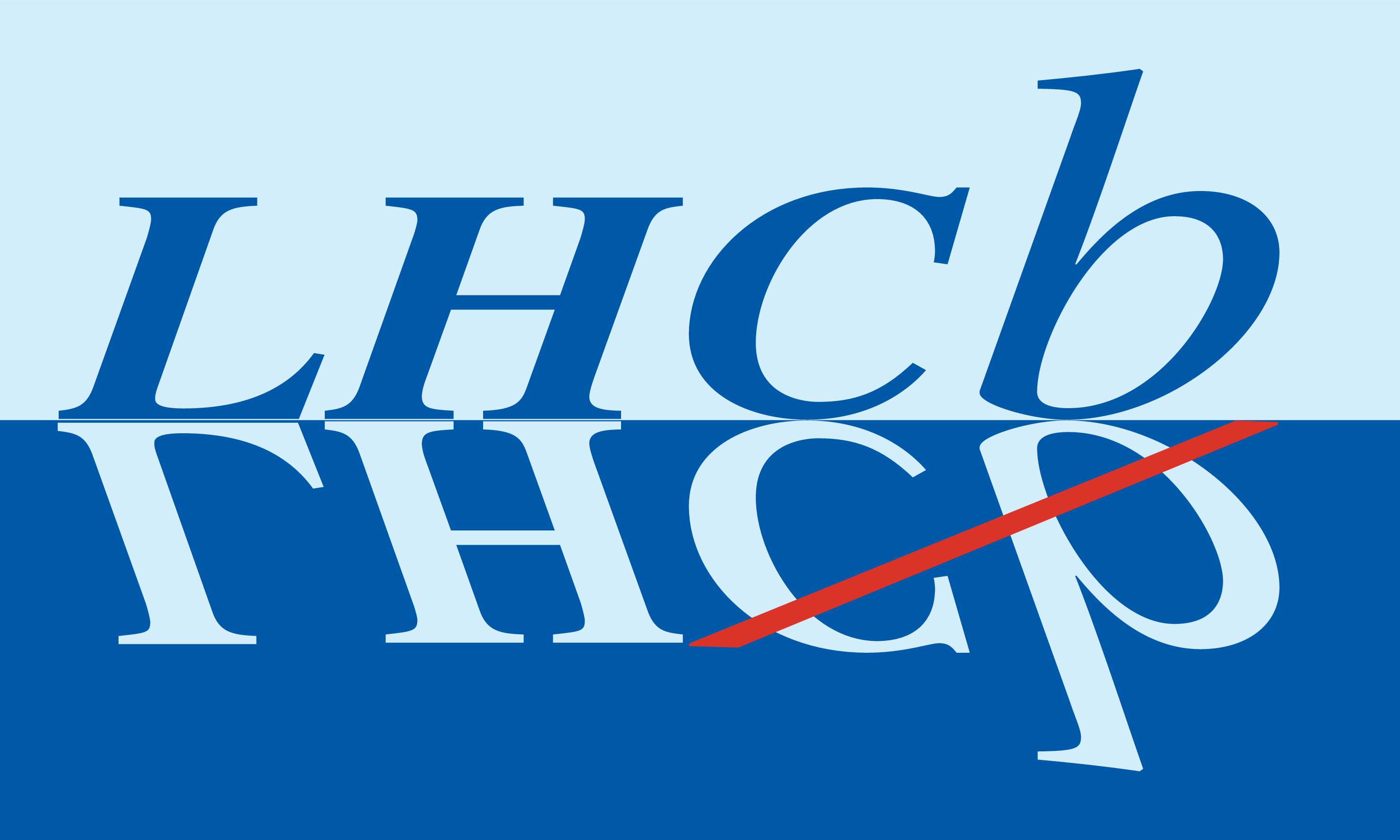}} & &}%
\\
 & & CERN-EP-2026-163 \\  
 & & LHCb-PAPER-2026-016 \\  
 & & 22 June 2026 \\ 
 & & \\
\end{tabular*}

\vspace*{4.0cm}

{\normalfont\bfseries\boldmath\huge
\begin{center}
  \papertitle 
\end{center}
}

\vspace*{2.0cm}

\begin{center}
\paperauthors\footnote{Authors are listed at the end of this paper.}
\end{center}

\vspace{\fill}

\begin{abstract}
  \noindent
  Differential measurements of the branching fraction and \CP asymmetry of the decay \Bpipmmumu are performed in intervals of the dimuon mass squared,~\qsq, using proton-proton collision data corresponding to an integrated luminosity of~9\invfb recorded by the \lhcb experiment at centre-of-mass energies of 7\tev,~8\tev and~13\tev. The ratios of the branching fractions of \Bpipmmumu and \BKpmmumu decays are also reported in the same \qsq intervals. The measured branching fractions are compatible with the predictions of the Standard Model, with consistency varying between 1.4$\sigma$ and 3.8$\sigma$ depending on the model assumptions.
\end{abstract}

\vspace*{2.0cm}

\begin{center}
  Submitted to
  JHEP
\end{center}

\vspace{\fill}

{\footnotesize 
\centerline{\copyright~\papercopyright. \href{\paperlicenceurl}{\paperlicence}.}}
\vspace*{2mm}

\end{titlepage}


\newpage
\setcounter{page}{2}
\mbox{~}
%
%
%
%


\renewcommand{\thefootnote}{\arabic{footnote}}
\setcounter{footnote}{0}


\cleardoublepage


\pagestyle{plain} 
\setcounter{page}{1}
\pagenumbering{arabic}


\section{Introduction}
\label{sec:Introduction}
 
Several measurements of processes involving flavour-changing neutral currents have provided tantalising hints of deviations from their Standard Model (SM) predictions. 
Branching fractions~\cite{LHCb-PAPER-2014-006,LHCb-PAPER-2015-023,LHCb-PAPER-2016-012,LHCb-PAPER-2021-014,BELLE:2019xld,Belle:2019oag} and angular observables~\cite{LHCb-PAPER-2025-041,LHCb-PAPER-2021-022,LHCb-PAPER-2020-041,ATLAS:2018gqc,CMS:2015bcy,CMS:2020oqb} associated with \bsmumu transitions show deviations from SM expectations across various final states. 
In addition, measurements of local and nonlocal amplitudes of \decay{B^\pm}{K^\pm \mu^+ \mu^-} decays present discrepancies with respect to the SM in the range between 2 to 4$\sigma$~\cite{LHCb-PAPER-2025-055}.
The origin of these tensions, whether they arise from contributions associated with physics beyond the Standard Model (BSM) or from underestimated hadronic effects, is the subject of ongoing and intense debate~\cite{Isidori_2025}.

Decays mediated by \bdmumu transitions offer a complementary set of observables with a significant suppression of the SM contribution by the Cabibbo-Kobayashi-Maskawa~(CKM) factor $\left|\Vtd/\Vts\right|^2$. 
This suppression gives sensitivity to BSM contributions of comparable magnitude to those required to account for the discrepancies seen in \bsmumu transitions, enabling stringent constraints on the flavour structure of possible BSM contributions. 
The decay \Bpipmmumu therefore constitutes a sensitive probe to search for BSM physics within the flavour sector.

The LHCb collaboration first observed \Bpipmmumu decays using a data sample corresponding to an integrated luminosity of 1\invfb~\cite{LHCb-PAPER-2012-020} of proton-proton ($pp$) collisions, and subsequently measured the differential branching fraction as a function of the dimuon invariant-mass squared, \qsq, with a data sample corresponding to an integrated luminosity of 3\invfb~\cite{LHCb-PAPER-2015-035}.
The results were found to be consistent with the predictions of the SM ~\cite{Ali:2013zfa,Hambrock:2015wka,FermilabLattice:2015cdh}, and include measurements of the \CP asymmetry integrated over \qsq, as well as of the ratio of branching fractions of the \Bpipmmumu and \BKpmmumu modes in the regions ${1< \qsq < 6\gevgevcccc}$ and ${15< \qsq < 22\gevgevcccc}$. From these observables, an estimate of $\left|\Vtd/\Vts\right|$ was derived.
Similarly, a recent angular analysis of \Bpipmmumu decays~\cite{LHCb-PAPER-2026-015} found consistency with the expectations of the SM in the considered \qsq regions.

In recent years there have been significant theoretical advances in understanding the \Bpipmmumu decay. Numerous studies have investigated its sensitivity to potential effects of BSM physics, as well as provided detailed computations of SM predictions for various observables across different \qsq~regions~\cite{Khodjamirian:2017fxg,Rusov:2019ixr,Biswas:2022lhu,Bause:2022rrs}.
It was recently claimed~\cite{kamenik2024cpoddwindowlongdistance} that a measurement of the \CP~asymmetry in \Bpipmmumu decays can be used to constrain the magnitude of the nonlocal effects in \bsmumu transitions under the assumption of SM and U-spin symmetries.

An improved method for extracting the ratio of CKM matrix elements $\left|\Vtd/\Vts\right|$ was proposed in Ref.~\cite{Khodjamirian:2017fxg}, using measurements of the ratio of branching fractions of \Bpipmmumu and \BKpmmumu decays, and the \CP~asymmetry of \Bpipmmumu decays simultaneously. A method inspired by this strategy has been implemented for this paper using the \texttt{GammaCombo} package~\cite{LHCb-PAPER-2016-032}, which provides a framework for fitting CKM elements with correlated inputs. It is used to extract $\left|\Vtd/\Vts\right|$ and the Wolfenstein parameters \cite{PhysRevLett.51.1945} $\rho$ and $\eta$.

This paper reports a measurement of the differential branching fraction $\dd{\cal B}/\dd\qsq$ and the \CP asymmetry \ACP of the \Bpipmmumu decay, and the ratio \BFratio of the branching fractions of \Bpipmmumu and \BKpmmumu decays, as well as their correlations, in intervals of \qsq. 
The analysis is performed using a data sample corresponding to an integrated luminosity of 9\invfb of $pp$ collisions collected by the \lhcb experiment at centre-of-mass energies of 7, 8 and 13\tev. The branching fractions are normalised using the control channel \BToJPsiK.
The observables associated with the measurement are defined in \qsq intervals as
\begin{eqnarray}
    \left<\frac{\dd\BF}{\dd\qsq}\right>_{\pi(K)^{\pm}} &\equiv& \frac{1}{q^2_{\textrm{max}} - {q^2_{\textrm{min}}}}\int_{q^2_{\textrm{min}}}^{q^2_{\textrm{max}}} \frac{\dd\BF (B^\pm \to \pi (K)^\pm \mu^+ \mu^-)}{\dd\qsq}\, \dd\qsq\,,\\
    \ACP &\equiv& \frac{\left<\frac{\dd\BF}{\dd\qsq}\right>_{\pi^-} - \left<\frac{\dd\BF}{\dd\qsq}\right>_{\pi^+}}{\left<\frac{\dd\BF}{\dd\qsq}\right>_{\pi^-} + \left<\frac{\dd\BF}{\dd\qsq}\right>_{\pi^+}}\,,\\
    \BFratio &\equiv& \frac{\left<\frac{\dd\BF}{\dd\qsq}\right>_{\pim}+\left<\frac{\dd\BF}{\dd\qsq}\right>_{\pip}}{\left<\frac{\dd\BF}{\dd\qsq}\right>_{\Km}+\left<\frac{\dd\BF}{\dd\qsq}\right>_{\Kp} }\,,
\end{eqnarray}
and are determined simultaneously from an extended unbinned maximum-likelihood fit to the data for all $\qsqmin<\qsq<\qsqmax$ intervals specified in Ref.~\cite{LHCb-PAPER-2015-035}. These \qsq intervals correspond to [0.1, 2.0], [2.0, 4.0], [4.0, 6.0], [6.0, 8.0], [11.0, 12.5], [15.0, 17.0], [17.0,~19.0], [19.0, 22.0] and [22.0, 25.0]\gevgevcccc, which excludes the \jpsi and \psitwos charmonia resonances. The \qsq resolution is significantly smaller than the widths of the \qsq intervals used in this analysis, and thus migration of events between \qsq intervals is negligible.
The highest \qsq interval is excluded for the kaon decay mode since it lies beyond the kinematic threshold accessible to this mode.
Consequently, \BFratio is not reported for this region. 
To reduce potential experimenter bias, the observables of the analysis were not examined until the entire experimental procedure had been fully finalised.

This paper is structured as follows. The detector and simulation are described in Sec.~\ref{sec:Detector}. The signal selection procedure is detailed in Sec.~\ref{sec:Selection}. The fitting procedure is presented in Sec.~\ref{sec:Fits}, and the evaluation of systematic uncertainties is discussed in Sec.~\ref{sec:Systematics}. The results of the analysis are reported in Sec.~\ref{sec:Results}, followed by concluding remarks in Sec.~\ref{sec:Conclusions}.

\section{Detector and simulation}
\label{sec:Detector}

The \lhcb detector~\cite{LHCb-DP-2008-001,LHCb-DP-2014-002} is a single-arm forward
spectrometer covering the \mbox{pseudorapidity} range $2<\eta <5$,
designed for the study of particles containing \bquark or \cquark
quarks. 
The detector used to collect the data analysed in this paper includes a high-precision tracking system consisting of a silicon-strip vertex detector surrounding the $pp$
interaction region, a large-area silicon-strip detector located
upstream of a dipole magnet with a bending power of about
$4\,{\mathrm{T\,m}}$, and three stations of silicon-strip detectors and straw
drift tubes placed downstream of the magnet.
The tracking system provides a measurement of the momentum, \ptot, of charged particles with
a relative uncertainty that varies from 0.5\% at low momentum to 1.0\% at 200\gevc.
The minimum distance of a track to a primary $pp$ collision vertex (PV), the impact parameter (IP), is measured with a resolution of $(15+29/\pt)\mum$,
where \pt is the component of the momentum transverse to the beam, in\,\gevc.
Different types of charged hadrons are distinguished using information
from two ring-imaging Cherenkov detectors. 
Photons, electrons and hadrons are identified by a calorimeter system consisting of scintillating-pad and preshower detectors, an electromagnetic and a hadronic calorimeter. Muons are identified by a system composed of alternating layers of iron and multiwire
proportional chambers.

The online event selection is performed by a trigger system,  which consists of a hardware stage~\cite{LHCb-DP-2012-004}, followed by a software stage, which applies a full event reconstruction.
For this analysis the hardware trigger retains events comprising at least one muon with high \pt, with the minimum threshold varying in the range 1200--2200\mevc depending on the data-taking period. 
In the subsequent software trigger, at least one of the final-state particles is required to have $\pt > 800\mevc$ and an IP greater than $100\mum$ with respect to all PVs in the event. A multivariate algorithm~\cite{BBDT,LHCb-PROC-2015-018} 
is then used to identify events containing secondary vertices formed by two or three tracks that are consistent with the decay of a \bquark~hadron. An algorithm is also used to select pairs of muons forming a vertex displaced from all PVs.
The data selected by the trigger system subsequently undergo a centralised, offline processing stage, designed to produce datasets of analysis-ready quality across the full \lhcb physics programme~\cite{Stripping}.

Simulation is used to model the effects of the detector acceptance and the imposed selection requirements.
In the simulation, $pp$ collisions are generated using
\pythia~\cite{Sjostrand:2007gs,*Sjostrand:2006za} 
with a specific \lhcb configuration~\cite{LHCb-PROC-2010-056}.
Decays of unstable particles
are described by \evtgen~\cite{Lange:2001uf}, in which final-state
radiation is generated using \photos~\cite{davidson2015photos}.
The signal \Bpipmmumu and \BKpmmumu decays are generated according to the model described in Ref.~\cite{Ali:1999mm} using form factors from Ref.~\cite{Ball:2004ye}.
The interaction of the generated particles with the detector, and its response,
are implemented using the \geant
toolkit~\cite{Allison:2006ve, *Agostinelli:2002hh} as described in
Ref.~\cite{LHCb-PROC-2011-006}. 

The simulated samples are corrected to account for known data-simulation differences in the \Bpm~production kinematics and detector occupancy. These correction factors are derived from control samples of \BToJPsiK decays, obtained by statistically subtracting the background contribution from the data~\cite{Pivk:2004ty}. Additional corrections are applied for track reconstruction, trigger and particle identification efficiencies taken from calibration samples~\cite{LHCb-DP-2013-002,LHCb-PUB-2014-039,LHCb-PUB-2016-021}.

The magnetic field deflects particles of opposite electric charge in opposite directions, which can induce detection asymmetries. 
Periodic reversal of the magnetic field polarity during data taking largely compensates for this effect. 
Small residual asymmetries in track reconstruction efficiency are taken into account by applying corrections derived from control data~\cite{LHCb-PAPER-2018-010}.

\section{Selection}
\label{sec:Selection}

The event selection closely follows that described in Ref.~\cite{LHCb-PAPER-2023-004} for the reconstruction of \Bpipmmumu candidates, which define the \textit{pion mode}. An alternative particle identification (PID) strategy is implemented to isolate \BKpmmumu candidates, thereby defining the \textit{kaon mode}. The pion mode sample is further divided into two subsamples according to the electric charge of the pion.

There are small, known imperfections in the modelling of the \Bpm-production kinematics and the observed track multiplicity.
Therefore a loose selection based on these variables is applied to ensure consistency between the data sample and the calibration samples used to correct the simulation. Each track is required to have a good track-fit quality, to lie within the geometrical and kinematic acceptance of the PID systems, and to have a significant IP with respect to all PVs in the event. Candidates formed of three-track combinations are required to be compatible with originating from a common decay vertex that is spatially displaced from every PV.

A subsequent selection is performed using a boosted decision tree (BDT) classifier~\cite{Breiman,AdaBoost}, as implemented in the TMVA toolkit~\cite{Hocker:2007ht,*TMVA4}, to separate signal from  combinatorial background.
Simulated \Bpipmmumu data are used as the signal proxy, while background is taken from the data sidebands above the signal peak in the reconstructed candidate mass regions \mbox{$5550 < m(\pipm\mup\mun) < 6000\mevcc$} and \mbox{$6500 < m(\pipm\mup\mun) < 7000$\mevcc}. 
The range \mbox{$6000 < m(\pipm\mup\mun) < 6500$\mevcc} is excluded to avoid potential contributions from resonant or nonresonant \decay{\Bcpm}{\pipm\mup\mun} decays~\cite{LHCb-PAPER-2023-037,LHCb-PAPER-2024-026}. 
To avoid potential training-related biases, a 10-fold~\cite{kFold} cross-validation procedure is used. The input variables to the BDT include transverse momenta and IP significances of the three candidate tracks, with additional variables described fully in Ref.~\cite{LHCb-PAPER-2023-004}.
To mitigate systematic effects, an identical requirement on the BDT output is applied to both the pion and kaon decay modes. The PID information is excluded from the BDT training procedure.

The requirement on the BDT output is optimised simultaneously with that on the pion PID to maximise the significance of the pion mode integrated across all \qsq intervals. The PID criteria are defined such that no candidates are shared between the pion and kaon-mode selections. A sample of \BToJPsiK events, used as the \textit{normalisation mode} for the measurement of the differential branching fraction, is extracted from the kaon-mode sample by limiting the invariant mass of the dimuon system to ${|m(\mup\mun) - m_\jpsi| < 50\mevcc}$, where $m_\jpsi$ is the known \jpsi~mass~\cite{PDG2024}.

Candidates are required to satisfy the condition $m(\Ph^\pm\mu^+\mu^-) > 5180\mevcc$, where $\Ph^\pm$ stands for a charged kaon or pion, to suppress partially reconstructed and semileptonic background contributions arising from decays of \Bd, \Bu and \Bs mesons into final states with a \pipm, \mup, \mun and additional unreconstructed particles. 
Candidates are excluded from the sample if they are consistent with originating from intermediate charmonium resonances under alternative mass-assignment hypotheses for the final-state particles.

\section{Fit strategy}
\label{sec:Fits}

After applying the selection criteria, the major contributions to the pion mode data sample are from the signal \Bpipmmumu decay itself, residual combinatorial background, and misidentified ${\BKpmmumu}$ decays in which the kaon is reconstructed under the pion mass hypothesis.
Minor contributions arise from hadronic \decay{\Bpm}{\pipm\Ph_1^\pm\Ph_2^\mp} decays, where the $\Ph_1^\pm \Ph_2^\mp$ system corresponds either to a dipion or a pion-kaon pair with both of these hadrons subsequently decaying into muons. Additional contributions arise from \Bs decays to \pip\pim\mup\mun final states, whose branching fractions are not well measured, in which a pion is not reconstructed. An example of such decays is \decay{\Bs}{f_0(980)\mu^+\mu^-}, with \decay{f_0(980)}{\pip\pim}. Any residual contribution from these sources, including semileptonic backgrounds that fall within the selected $m(\pi\mu\mu)$ mass region, can be accommodated by the \decay{\Bs}{\pip \pim \mup\mun} component and is covered by the partial-reconstruction shape systematic uncertainty.
For the kaon- and normalisation-mode samples, the only contributions taken into account are the \BKpmmumu and \BToJPsiK decays, and combinatorial backgrounds.

To extract \diffBF, \ACPRaw, and \BFratio, an extended unbinned maximum-likelihood fit is performed to the data using the \roofit framework~\cite{roofit}. 
The fit is performed simultaneously to the ${\Bpipmumu}$, ${\Bpimmumu}$ and ${\BKpmmumu}$ data samples, parameterising the yields of the components in each \qsq bin as
\begin{eqnarray}
    N_{\text{\Bpipmumu}} &=& \dfrac{1-\ACPRaw}{2}\left<\frac{\dd\BF}{\dd\qsq}\right>_{\pi^\pm}\Delta\qsq N_{B} \, \varepsilon_\text{\Bpipmumu}\,,\\    
    N_{\text{\Bpimmumu}} &=& \dfrac{1+\ACPRaw}{2} \left<\frac{\dd\BF}{\dd\qsq}\right>_{\pi^\pm} \Delta\qsq N_{B} \,\varepsilon_\text{\Bpimmumu}\,,\\
    N_{\text{\BKpmmumu}} &=& \frac{1}{R_{\pi/K}} \left<\frac{\dd\BF}{\dd\qsq}\right>_{\pi^\pm} \Delta\qsq N_{B}\,\varepsilon_\text{\BKpmmumu}\,,
\end{eqnarray}
where
\begin{eqnarray}
    N_{B} &\equiv& \frac{N_\text{\BToJPsiK}}{\BF(\BToJPsiK)\; \varepsilon_\text{\BToJPsiK}}\,,
\end{eqnarray}
$N_X$ and $\varepsilon_X$ denote the yields and total efficiencies, respectively, for the decay modes $X$ indicated in the subscripts, and $\Delta\qsq$ denotes the \qsq interval width. The raw \CP asymmetry obtained from the fit is denoted \ACPRaw and corresponds to the asymmetry of the efficiency-corrected \Bpipmmumu yields. The selection efficiencies for each decay mode are determined from simulation with corrections based on the data applied. For the pion mode, they are measured separately for the two charges thus accounting for PID selection asymmetries. The \CP asymmetry of the kaon mode is assumed to be zero, consistent with previous measurements~\cite{LHCb-PAPER-2013-043}. For the normalisation mode, the quantity $\varepsilon_\text{\BToJPsiK}$ is evaluated using the combined \decay{\Bp}{\jpsi\Kp} and \decay{\Bm}{\jpsi\Km} simulation samples.
The yield of the \BToJPsiK mode is fixed to the value obtained from an independent fit to the combined \decay{\Bp}{\jpsi\Kp} and \decay{\Bm}{\jpsi\Km} normalisation-mode data. The normalisation branching fraction ${\cal B}(\BToJPsiK)$ incorporates the known branching fraction of the $\jpsi\to\mup\mun$ decay and is given by ${(6.08\pm0.12)\times10^{-5}}$~\cite{PDG2024}.

The yield of the misidentified kaon component is obtained by scaling the yield of the correctly identified kaon component. The scaling factor, which is the relative efficiency for reconstructing the decay as a pion rather than as a kaon, is corrected for production and detection asymmetries using the raw ${\ACPRaw(\BToJPsiK)}$ value determined from the normalisation-mode fit.

The \Bpipmmumu and \BKpmmumu components in their respective fits are described using a modified hyperbolic function with power-law tails~\cite{Santos:2013gra}. 
The shape parameters of this function are allowed to vary in the fit to data within Gaussian constraints derived from the normalisation-mode fit. 
Differences in mass resolution are accounted for by scaling the width of \BToJPsiK decays by a factor obtained from simulation in each \qsq\ interval. 

The combinatorial background is modelled with a linear function, which is validated using a sample to which the same selection criteria are applied, with the exception of an alternative requirement that the two muons have identical electric charge.
For the normalisation-mode fit, the combinatorial background is modelled with an exponential function due to the high statistical precision of the fit sample.

The misidentified \BKpmmumu component is modelled with a Gaussian function with power-law tails on both sides of the distribution~\cite{Skwarnicki:1986xj}.
The model shape parameters are shared in a simultaneous fit to the selected \BKpmmumu events, reconstructed with the final-state hadron mass assumed to be a pion, and weighted to account for differences in the PID selection efficiencies between the pion and kaon modes. 

Contamination arising from ${\Bpm\to \pipm h^\pm h^\mp}$ decays is modelled using the sum of two Gaussian functions, each with a power-law tail, one on the low-mass side and one on the high-mass side of the peak. 
The shape parameters of these functions are fixed from simulated \BuKPiPi and \BuPiPiPi decays, weighted to reproduce the Dalitz-plane structure measured by BaBar~\cite{BaBar:2008lpx} and LHCb~\cite{LHCb-PAPER-2019-017}, respectively.
The corresponding yields are allowed to vary in the fit to data within Gaussian constraints, which are derived from control regions of the data obtained by relaxing the muon-identification requirements to ${\Bpm\to \pipm h^\pm h^\mp}$ data. 
In these control regions, muon-identification criteria are imposed on a single hadron at a time, and the resulting misidentification rates are then extrapolated to estimate how often two hadrons are misidentified simultaneously.
Correlations between the two hadrons are taken into account by applying correction factors derived from calibration samples of reconstructed \KS\to\pip\pim and \Dz\to\Km\pip decays~\cite{LHCb-PUB-2016-021}.

\begin{figure}[!tb]
  \centering
    \includegraphics[page=1, width=0.49\linewidth]{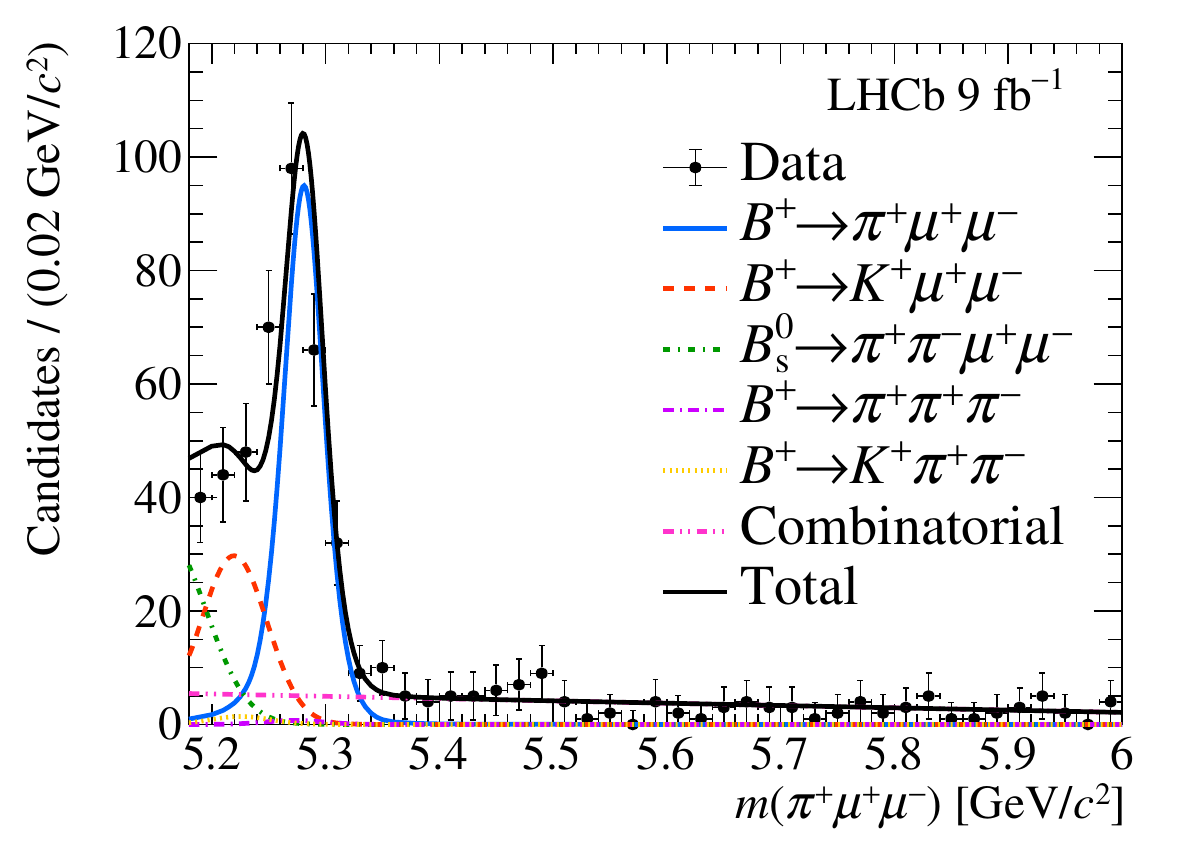}
    \includegraphics[page=1, width=0.49\linewidth]{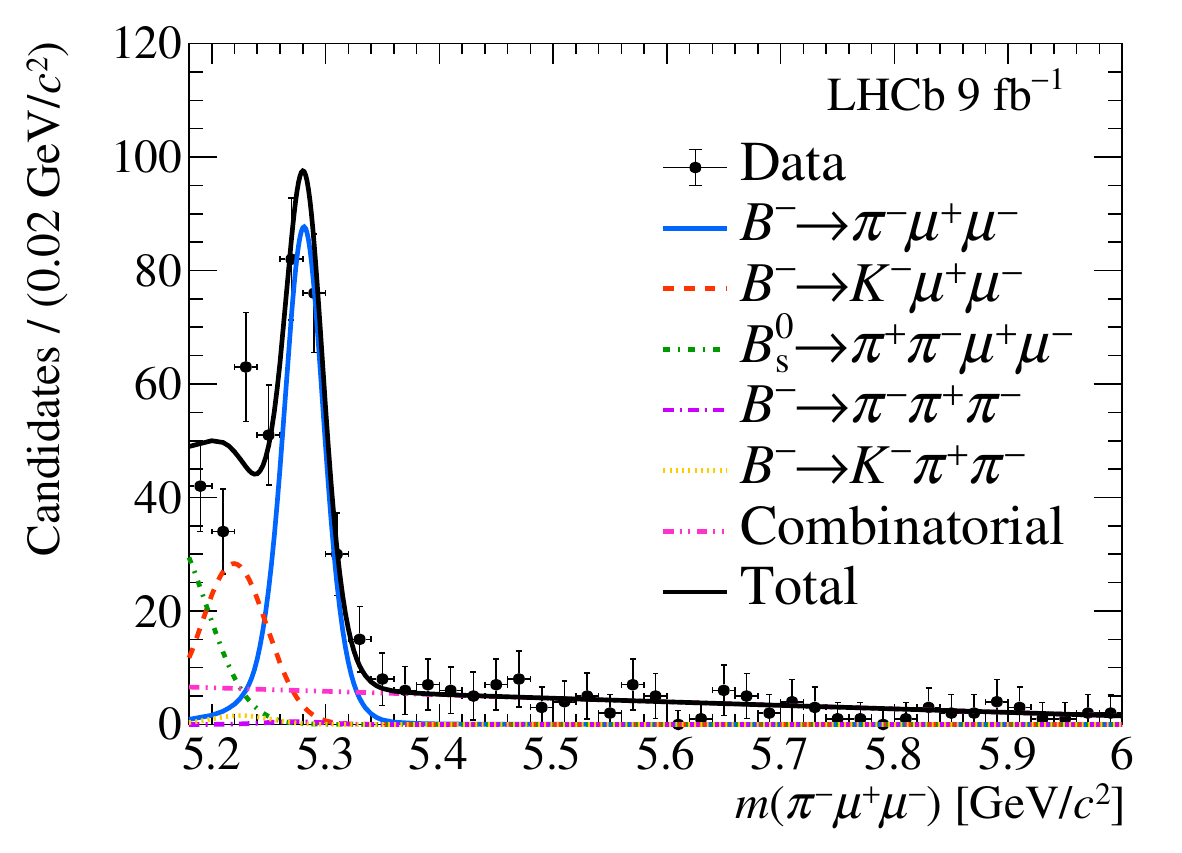}
    \includegraphics[page=1, width=0.49\linewidth]{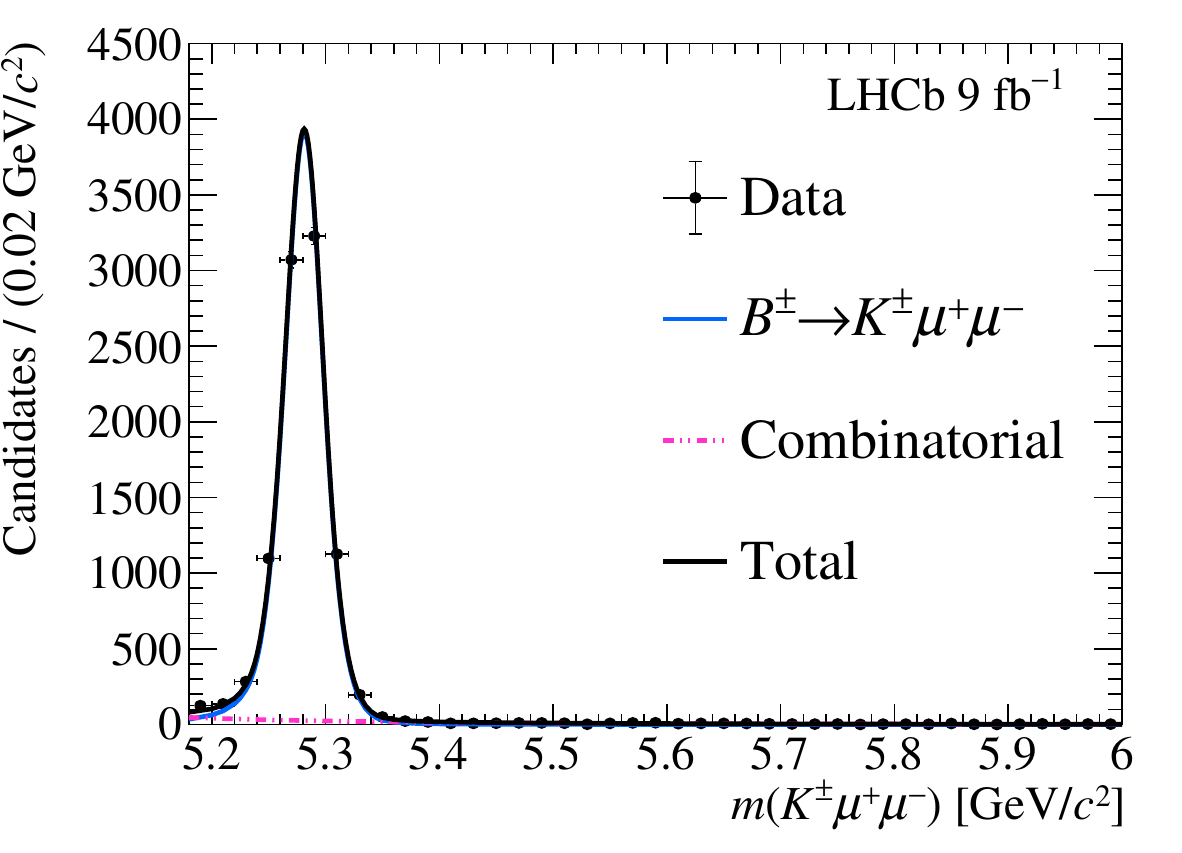}
  \caption{
    Distributions of the reconstructed mass for (top left) \Bpipmumu and (top right) \Bpimmumu candidates, and (bottom) \BKpmmumu candidates, combined over all \qsq intervals.
    The result of the fits described in the text are also shown.
  }
  \label{fig:fitprojections}
\end{figure}

The \decay{\Bs}{\pip \pim \mup\mun} component is modelled using a Gaussian function with an exponential tail on the low-mass side of the peak. 
The shape parameters are fixed from simulation and the yield allowed to vary freely in the fit to data.
Figure~\ref{fig:fitprojections} shows the $\pi^+\mu^+\mu^-$, $\pi^-\mu^+\mu^-$, and $K^\pm\mu^+\mu^-$ invariant-mass distributions, with the fit overlaid, combined over all \qsq intervals.

Uncertainties on the observables are allowed to be asymmetric and are determined from the likelihood function using the \textsc{MINOS} procedure~\cite{Eadie:100342}. The statistical correlation matrix is obtained from the Hessian matrix of the negative log-likelihood evaluated at its minimum. 

The validity and robustness of the fit are evaluated using ensembles of a large number of pseudoexperiments generated from the models described in this section, with shape parameters obtained from the fit to data.
The true values of the observables in these pseudoexperiments are drawn from a broad range around the SM predictions. 
For the majority of observables, the pull distributions indicate negligible bias and correct statistical coverage.
Consequently, no statistical bias corrections are applied to the observables.

The raw \ACPRaw observable obtained from the fit requires a correction for possible asymmetry in \Bpm production, \AProd, and from detection asymmetry effects, \ADet, such that
\begin{equation}
    \ACP = \ACPRaw - \AProd - \ADet\,,
\end{equation}
which is a valid approximation under the assumption that the production and detection asymmetries are small. The production asymmetry~\cite{LHCb-PAPER-2021-049} is determined from the measured values of \AProd reported in Refs.~\cite{LHCb-PAPER-2016-062,LHCb-PAPER-2023-007}, corrected using the simulated \Bpm~kinematic distributions in each \qsq interval, and is found to lie between $-0.5$\% and $-0.7$\%. 
Charge asymmetries in the PID selection are incorporated in the efficiency determinations, while charge asymmetries in the tracking efficiency for pions are corrected using calibration tables following Ref.~\cite{LHCb-PAPER-2018-010}, taking into account the pion kinematics in each \qsq interval. Since Ref.~\cite{LHCb-PAPER-2018-010} was derived using 2011--2012 data, the uncertainties corresponding to these corrections are increased for the 2015--2018 sample to account for possible variations between data-taking periods.
The resulting values for \ADet vary between $-0.16$\% and $-0.25$\%. Cross-checks are performed by splitting the dataset according to the magnet polarity and consistent results are obtained.

\section{Systematic uncertainties}
\label{sec:Systematics}

Systematic uncertainties can be classified into two categories: those originating from uncertainties in the model shapes used in the fits and those arising from uncertainties in the efficiency determinations. Table~\ref{tab:syst_breakdown_sources} shows a list of systematic sources considered in this analysis.
Uncertainties associated with the model shapes are estimated by generating pseudodata from alternative models, using the shape variations discussed below, and subsequently performing fits to these pseudodata samples with both the nominal and alternative models. The mean difference between the observables obtained from the two fits is taken as the systematic uncertainty assigned to these observables. 

The systematic uncertainty associated with the \Bpipmmumu signal mass shape is evaluated by adding an additional Gaussian function to the nominal model. The parameters of this alternative shape are determined from simulation. In addition, systematic uncertainties associated with the $B^\pm \to h^\pm \mu^+ \mu^-$ signal widths, obtained by scaling the nominal \Bpipmmumu and \BKpmmumu lineshape widths relative to the \BToJPsiK width, are determined from simulation.
For the combinatorial background shape, a systematic uncertainty is assigned by adopting an alternative parameterisation based on an exponential function. 
For the ${\Bpm\to \pipm h^\pm h^\mp}$ contribution, the corresponding systematic uncertainty is evaluated by simultaneously varying both the shape parameters and the yields within their respective uncertainties.  

For the \decay{\Bs}{\pip\pim\mup\mun} contribution, a systematic uncertainty is determined by employing an alternative shape derived from a kernel density estimator.
An additional source of systematic uncertainty is evaluated to account for any potential leakage from partially reconstructed \decay{B^{0,\pm}}{\Kstar(892)^{0,\pm}\mup\mun} decays. 
This contribution originates from events where the pion from the $\decay{\Kstar(892)^{0,\pm}}{\Kpm\Ppi^{\mp,0}}$ decay is not reconstructed and the charged kaon is misidentified as a pion. 
The impact of this background is quantified by incorporating an additional component into the pseudodata generation procedure. 
The shape of this component is derived from the Gaussian tail of a fit to simulated \decay{\Bz}{\Kstar(892)^{0}\mup\mun} decays, while its yield is fixed according to the measured branching fractions~\cite{PDG2024}.
In the case of the misidentified \BKpmmumu contribution, the nominal model describes the distribution observed in the \BKpmmumu sample with high statistical precision. 
As a result, the only systematic uncertainty attributed to this component is that associated with the efficiency correction. 

Systematic uncertainties in the efficiency determinations are evaluated using ensembles of pseudoexperiments, following an analogous procedure to that employed for the model shape uncertainties. Pseudodata are generated by varying the efficiency parameters and subsequently fitting them with the nominal values. These fluctuations are implemented coherently across all decay modes to correctly account for correlations.
Uncertainties associated with the kinematic and occupancy corrections are estimated by removing them and assessing their impact.
Track reconstruction, trigger, and PID efficiency corrections are varied within their statistical uncertainties.

The simulated samples used for the efficiency determination are generated according to the SM predictions~\cite{Ali:1999mm,Ball:2004ye}. Since the efficiency is evaluated in intervals of \qsq, there is an additional uncertainty associated with the true \qsq distribution within each interval, which is sensitive to theoretical uncertainties in the underlying model used to generate the simulation. To quantify the systematic uncertainty associated with the simulation physics model, the \texttt{flavio} package~\cite{straub:2018flavio} is used to generate a set of event weights for the simulation by fluctuating the default model parameters according to their covariance matrix.

The differential branching fraction is subject to systematic uncertainties arising from knowledge of the normalisation mode branching fraction \cite{PDG2024}.
The \ACP observables are subject to additional systematic uncertainties arising from production and detection asymmetries. These contributions are quantified by propagating to the measured values both statistical and systematic uncertainties associated with the corresponding corrections.

Table~\ref{tab:syst_breakdown_sources} shows the ranges of systematic uncertainties across all \qsq bins for each source. The ratio between the total systematic and statistical uncertainties are evaluated in each \qsq interval, and are in the ranges 0.14--0.30 for~\diffBF, 0.07--0.17 for~\ACP, and~0.12--0.25 for~\BFratio.
For the branching-fraction measurements, the dominant source of systematic uncertainty arises from the uncertainty on the normalisation branching fraction. The next largest contributions (and largest for \ACP and \BFratio) come from the modelling of the signal, combinatorial, and partially reconstructed background shapes. The complete covariance matrices are provided on the HEPData repository~\cite{HEPdata}.

\renewcommand{\arraystretch}{1.2}
\begin{table}[]
\centering
\caption{Systematic uncertainties across all \qsq bins for each systematic source, where the lowest and highest values are shown.}
\label{tab:syst_breakdown_sources}
\begin{tabular}{l|c|c|c} \hline
Source & $\diffBF \times 10^{10}$ & $\ACP \times 10^2$ & $\BFratio \times 10^{3}$ \\ \hline
$B^\pm \to \pi^\pm \mu^+ \mu^-$ signal shape & $ 0.02-0.23 $ & $ 0.11-1.95 $ & $ 0.08-0.82 $ \\
$B^\pm \to h^\pm \mu^+ \mu^-$ signal widths & $ 0.00-0.08 $ & $ 0.01-0.08 $ & $ 0.05-0.28 $ \\
Combinatorial background shape & $ 0.00-0.20 $ & $ 0.11-1.63 $ & $ 0.00-0.97 $  \\
$B^\pm \to \pi^\pm h^\pm h^\mp$ backgrounds & $ 0.00-0.02 $ & $ 0.01-0.38 $ & $ 0.01-0.11 $ \\
$B^0_s \to \pip \pim \mu^+ \mu^-$ shape & $ 0.01-0.25 $ & $ 0.24-2.83 $ & $ 0.12-1.45 $ \\
\decay{B^{0,\pm}}{\Kstar(892)^{0,\pm}\mup\mun} background & $ 0.00-0.01 $ & $ 0.00-0.05$ & $ 0.00-0.04 $ \\
Kinematic and occupancy corrections & $ 0.00-0.09 $ & $ 0.05-0.44 $ & $ 0.04-1.00 $ \\
Tracking efficiency & $ 0.02-0.10 $ & $ 0.00-0.01$ & $ 0.13-0.42 $  \\
Trigger efficiency & $ 0.02-0.23 $ & $ 0.00-0.01$ & $ 0.22-0.72 $ \\
PID efficiency & $ 0.00-0.06 $ & $ 0.00-0.01$ & $ 0.06-0.28 $ \\
Simulation physics model & $ 0.00-0.01 $ & $ 0.00-0.01$ & $ 0.03-0.10 $ \\
Normalisation mode &  $ 0.10-0.54 $ & -- & -- \\
$\mathcal{A}^{\CP}$ corrections & -- & $ 0.31-0.35 $ & -- \\ \hline
Total &  $ 0.12-0.72 $ & $ 0.75-3.45 $ & $ 0.61-2.31 $  \\ \hline
\end{tabular}
\end{table}
\section{Results}
\label{sec:Results}

\begin{figure}[!tb]
  \centering
    \includegraphics[width=0.49\linewidth]{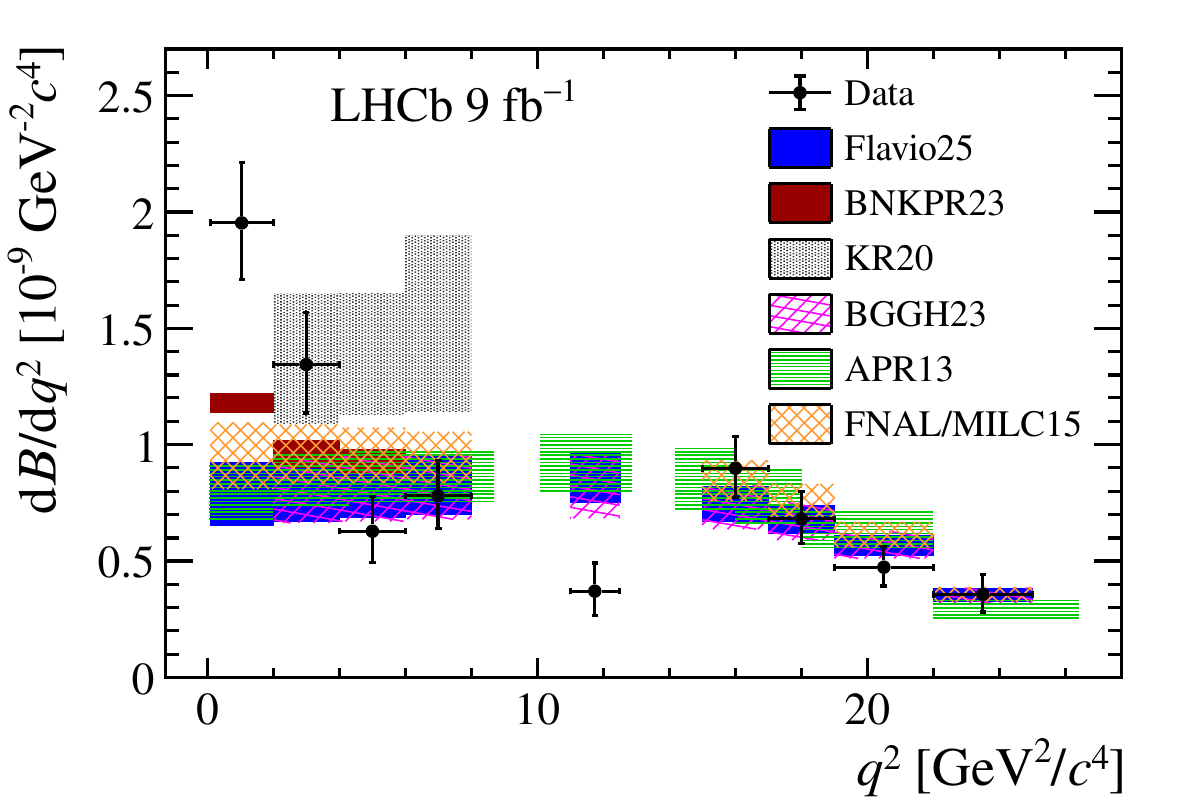}
    \includegraphics[width=0.49\linewidth]{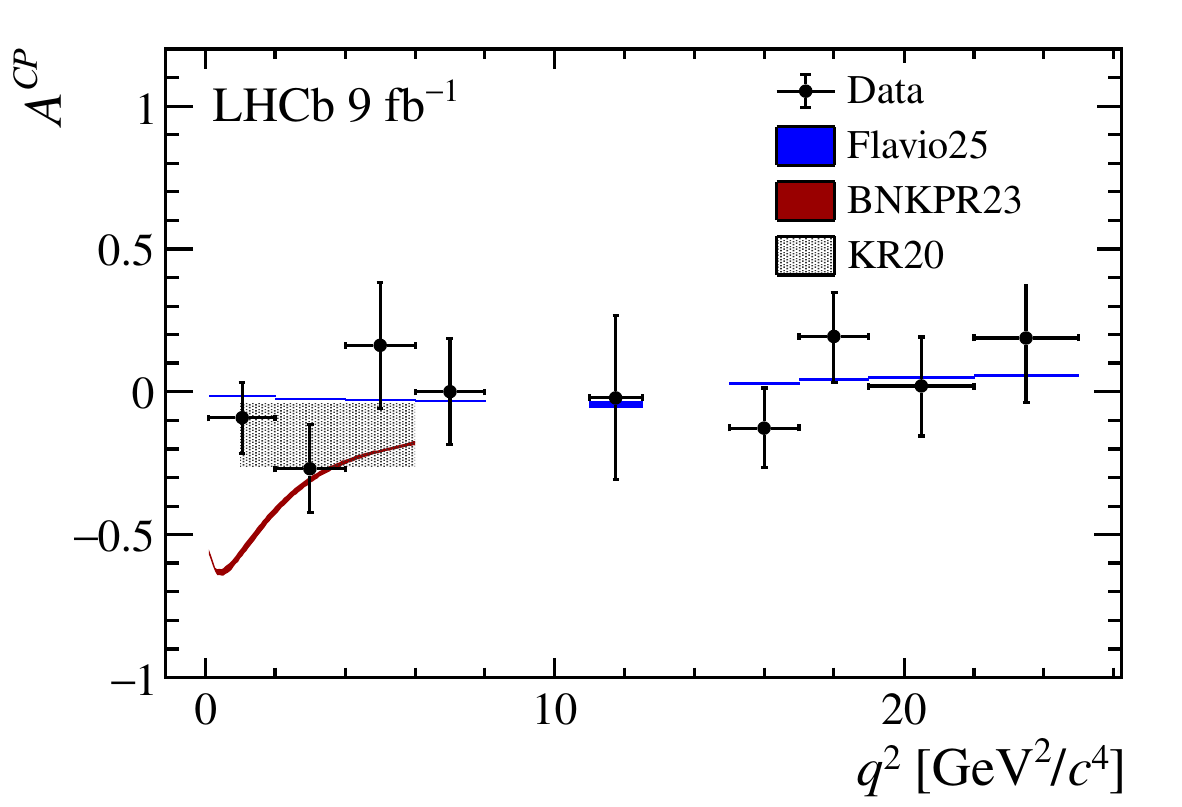}
    \includegraphics[width=0.49\linewidth]{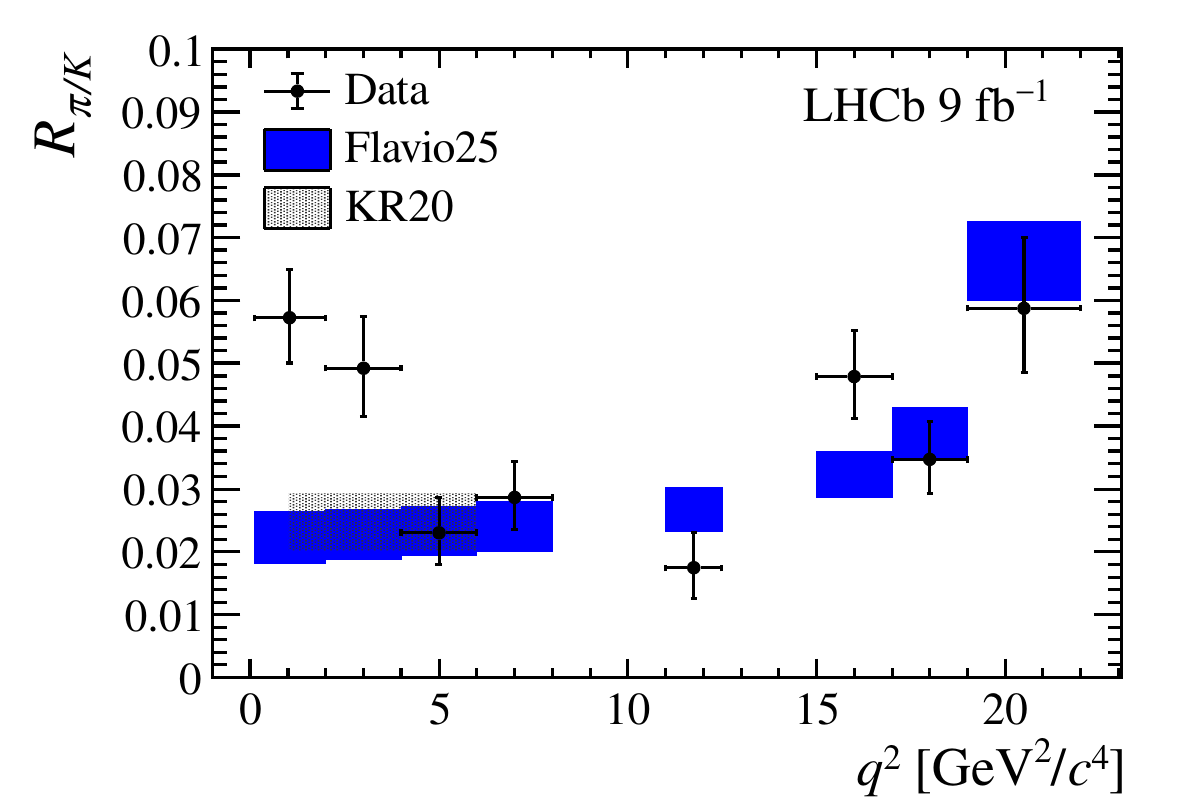}
  \caption{
    Measured values for (top left) the differential branching fraction of the \Bpipmmumu decay, (top right) the \CP asymmetry in the \Bpipmmumu channel, and (bottom) the ratio of differential branching fractions between the \Bpipmmumu and \BKpmmumu modes, all evaluated in bins of the dimuon invariant-mass squared, \qsq. 
    The uncertainties represent the quadratic sum of statistical and systematic contributions.
    The results are compared with SM predictions~\cite{straub:2018flavio,Biswas:2022lhu,Rusov:2019ixr,Khodjamirian:2017fxg,Bause:2022rrs,Ali:2013zfa,FermilabLattice:2015cdh}.
  }
  \label{fig:results}
\end{figure}

The complete set of results for the observables is shown in Fig.~\ref{fig:results}, compared with SM predictions~\cite{straub:2018flavio,Biswas:2022lhu,Rusov:2019ixr,Khodjamirian:2017fxg,Bause:2022rrs,Ali:2013zfa,FermilabLattice:2015cdh}. 
The numerical central values, statistical uncertainties and their correlations are listed in the Appendix.

A $\chi^2$ test is used to compute the compatibility between the SM theory predictions and the measured differential branching fraction. This test is performed using predictions whose \qsq regions are aligned to those used in this analysis, as presented in Table~\ref{tab:chisqs}. 
The compatibility with the predictions considered varies between 1.4 and 3.8$\sigma$, with the largest deviations seen with the Flavio25 predictions, which do not model the light-quark and charmonia resonances. The smallest deviations are seen for BGGH23~\cite{Bause:2022rrs} where the $0.1 < q^2 < 2.0$\gevgevcccc interval is excluded.

\begin{table}[]
    \centering
    \caption{Compatibility, evaluated using a $\chi^2$ test, between SM theoretical predictions and measured differential branching fraction in each \qsq bin, where the \qsq regions are aligned.}
    \label{tab:chisqs}
    \begin{tabular}{c|c|c}
    Prediction & \qsq range [\gevgevcccc] & Compatibility ($\sigma$) \\ \hline 
      BGGH23~\cite{Bause:2022rrs}  & $[2,8] \cup [11,12.5] \cup [15,25]$ & $1.35$ \\
      KR20~\cite{Rusov:2019ixr, Khodjamirian:2017fxg}  & $[2,8]$ & $2.03$ \\
      FNAL/MILC15~\cite{FermilabLattice:2015cdh} & $[0.1,8] \cup [15,25]$ & $2.45$ \\
      BNKPR23~\cite{Biswas:2022lhu} & $[0.1,6]$ &  $3.14$ \\
      Flavio25~\cite{straub:2018flavio} & $[0.1,8] \cup [11,12.5] \cup [15,25]$ & $3.78$
    \end{tabular}
\end{table}

The ratio of CKM matrix elements $\left|\Vtd/\Vts\right|$ is determined using \BFratio and \ACP in the region ${2.0<\qsq<6.0}$\gevgevcccc. Following the methodology described in Ref.~\cite{Khodjamirian:2017fxg}, as implemented in the \texttt{GammaCombo} package~\cite{GammaCombo}, this yields $\left|\Vtd/\Vts\right|=0.241^{+0.030}_{-0.027}$ which is $1.3\sigma$ above the world average \cite{PDG2024}. In this determination, the Wolfenstein parameters $\rho$ and $\eta$ are treated as nuisance parameters. Theoretical input parameters, which account for nonlocal contributions, and correlations among them are included using values provided by the authors of Ref.\cite{Khodjamirian:2017fxg}.

In addition, a log-likelihood profile as a function of the Wilson coefficient \C9 is constructed using the \texttt{flavio} package~\cite{straub:2018flavio}. This approach is motivated by the observation that the \bsll anomalies can be interpreted by a shift in the Wilson coefficient \C9 \cite{Gubernari:2022hxn, Capdevila:2023yhq}. The default theoretical input parameters are used, which do not include light-quark contributions, and theoretical uncertainties are included via the fast likelihood implementation provided by \texttt{flavio}. The profile is obtained from the binned branching-fraction data in the range ${0.1<\qsq<8.0\gevgevcccc}$. 
Based on Wilks' theorem~\cite{Wilks:1938dza}, \C9 is found to be compatible with the SM expectation at the level of $0.6\sigma$, assuming all other Wilson coefficients take their SM values. At the same time, assuming minimal flavour violation and the default predictions in \texttt{flavio}~\cite{straub:2018flavio}, this value for $\mathcal{C}_9^{bd\mu\mu}$ is consistent within $1.7\sigma$ with the value of $\mathcal{C}_9^{bs\mu\mu}$ required to explain the anomalies seen in the decay $B^0\to K^{*0}\mu^+ \mu^-$~\cite{LHCb-PAPER-2025-041}.

\section{Conclusions}
\label{sec:Conclusions}

The differential branching fraction of the \Bpipmmumu decay, the associated \CP asymmetry, and the ratio of the differential branching fractions of the \Bpipmmumu and \BKpmmumu decays are measured using $pp$ collision data corresponding to an integrated luminosity of 9\invfb collected with the LHCb detector.  
The measurements are performed in intervals of the dimuon invariant-mass squared, \qsq.
The \CP asymmetry and the branching-fraction ratio are measured for the first time for various \qsq intervals.
The results for the differential branching fraction are found to be consistent with, and supersede, the previous LHCb measurement~\cite{LHCb-PAPER-2015-035}.
Some localised deviations from the SM predictions are seen up to the level of three standard deviations.
The compatibility with the different SM predictions in the differential branching fraction varies between 1.4$\sigma$ and~3.8$\sigma$, with the largest deviations seen for the Flavio25 predictions that do not model the effect of light-quark and charmonia resonances.

\section*{Acknowledgements}
%
%
\noindent We express our gratitude to our colleagues in the CERN
accelerator departments for the excellent performance of the LHC. We
thank the technical and administrative staff at the LHCb
institutes.
We acknowledge support from CERN and from the national agencies:
ARC (Australia);
CAPES, CNPq, FAPERJ and FINEP (Brazil); 
MOST and NSFC (China); 
CNRS/IN2P3 and CEA (France);  
BMFTR, DFG and MPG (Germany);
INFN (Italy); 
NWO (Netherlands); 
MNiSW and NCN (Poland); 
MEC/IFA (Romania); 
MICIU and AEI (Spain);
SNSF and SER (Switzerland); 
NASU (Ukraine); 
STFC (United Kingdom); 
DOE NP and NSF (USA).
We acknowledge the computing resources that are provided by ARDC (Australia), 
CBPF (Brazil),
CERN, 
IHEP and LZU (China),
IN2P3 (France), 
KIT and DESY (Germany), 
INFN (Italy), 
SURF (Netherlands),
Polish WLCG (Poland),
IFIN-HH (Romania), 
PIC (Spain), CSCS (Switzerland), 
GridPP (United Kingdom),
and NSF (USA).  
We are indebted to the communities behind the multiple open-source
software packages on which we depend.
Individual groups or members have received support from
RTP (Australia), 
FWO Odysseus grant G0ASD25N (Belgium), 
Key Research Program of Frontier Sciences of CAS, CAS PIFI, CAS CCEPP (China); 
Minciencias (Colombia);
EPLANET, Marie Sk\l{}odowska-Curie Actions, ERC and NextGenerationEU (European Union);
A*MIDEX, ANR, IPhU and Labex P2IO, and R\'{e}gion Auvergne-Rh\^{o}ne-Alpes (France);
Alexander-von-Humboldt Foundation (Germany);
ICSC (Italy); 
Severo Ochoa and Mar\'ia de Maeztu Units of Excellence, GVA, XuntaGal, GENCAT, InTalent-Inditex and Prog.~Atracci\'on Talento CM (Spain);
the Leverhulme Trust, the Royal Society and UKRI (United Kingdom).


\clearpage
\section*{Appendix}
\label{sec:appendix}
\appendix


Table~\ref{tab:bintest_stats} shows the central values, statistical uncertainties and statistical correlations of all observables in all \qsq bins. Note the large correlation between \diffBF and \BFratio arises from the fact that both observables depend on the pion mode yield. 
\begin{table}[!b]
\centering
\caption{The central values, statistical uncertainties and corresponding correlation matrices of all observables in all \qsq intervals. These must be combined with the systematic covariance matrix between all bins available in Ref.~\cite{HEPdata}.}
\begin{tabular}{ll|c|cccc}
\multirow{2}{*}[-2.5ex]{$q^2$ Region [\gevgevcccc]} &
\multirow{2}{*}[-2.5ex]{Observable} &
\multirow{2}{*}[-2.5ex]{Value} &
\multicolumn{3}{c}{Statistical Correlation} \\
 & & &
{\diffBF} &
{$\ACP$} &
{\BFratio} \\\hline
\multirow{3}{0.19\textwidth}{$0.1 < q^2 < 2.0$} & \diffBF $\times 10^9$ & $\phantom{-}1.95^{+0.25}_{-0.23}$ & $\phantom{-}1.00$ & $\phantom{-}0.09$ & $\phantom{-}0.98$ & \\
 & \ACP & $-0.09^{+0.12}_{-0.12}$ & $\phantom{-}0.09$ & $\phantom{-}1.00$ & $\phantom{-}0.09$ \\
 & \BFratio $\times 10^2$ & $\phantom{-}5.72^{+0.76}_{-0.70}$ & $\phantom{-}0.98$ & $\phantom{-}0.09$ & $\phantom{-}1.00$ \\
\hline
\multirow{3}{0.19\textwidth}{$2.0 < q^2 < 4.0$} & \diffBF $\times 10^9$ & $\phantom{-}1.34^{+0.22}_{-0.20}$ & $\phantom{-}1.00$ & $\phantom{-}0.09$ & $\phantom{-}0.98$ & \\
 & \ACP & $-0.27^{+0.15}_{-0.15}$ & $\phantom{-}0.09$ & $\phantom{-}1.00$ & $\phantom{-}0.09$ \\
 & \BFratio $\times 10^2$ & $\phantom{-}4.92^{+0.82}_{-0.75}$ & $\phantom{-}0.98$ & $\phantom{-}0.09$ & $\phantom{-}1.00$ \\
\hline
\multirow{3}{0.19\textwidth}{$4.0 < q^2 < 6.0$} & \diffBF $\times 10^9$ & $\phantom{-}0.63^{+0.14}_{-0.13}$ & $\phantom{-}1.00$ & $-0.13$ & $\phantom{-}0.99$ & \\
 & \ACP & $\phantom{-}0.16^{+0.22}_{-0.22}$ & $-0.13$ & $\phantom{-}1.00$ & $-0.13$ \\
 & \BFratio $\times 10^2$ & $\phantom{-}2.30^{+0.54}_{-0.48}$ & $\phantom{-}0.99$ & $-0.13$ & $\phantom{-}1.00$ \\
\hline
\multirow{3}{0.19\textwidth}{$6.0 < q^2 < 8.0$} & \diffBF $\times 10^9$ & $\phantom{-}0.78^{+0.15}_{-0.14}$ & $\phantom{-}1.00$ & $-0.02$ & $\phantom{-}0.99$ & \\
 & \ACP & $\phantom{-}0.00^{+0.19}_{-0.18}$ & $-0.02$ & $\phantom{-}1.00$ & $-0.02$ \\
 & \BFratio $\times 10^2$ & $\phantom{-}2.87^{+0.57}_{-0.52}$ & $\phantom{-}0.99$ & $-0.02$ & $\phantom{-}1.00$ \\
\hline
\multirow{3}{0.19\textwidth}{$11.0 < q^2 < 12.5$} & \diffBF $\times 10^9$ & $\phantom{-}0.37^{+0.12}_{-0.10}$ & $\phantom{-}1.00$ & $-0.06$ & $\phantom{-}0.99$ & \\
 & \ACP & $-0.02^{+0.29}_{-0.29}$ & $-0.06$ & $\phantom{-}1.00$ & $-0.06$ \\
 & \BFratio $\times 10^2$ & $\phantom{-}1.74^{+0.55}_{-0.47}$ & $\phantom{-}0.99$ & $-0.06$ & $\phantom{-}1.00$ \\
\hline
\multirow{3}{0.19\textwidth}{$15.0 < q^2 < 17.0$} & \diffBF $\times 10^9$ & $\phantom{-}0.90^{+0.13}_{-0.12}$ & $\phantom{-}1.00$ & $\phantom{-}0.00$ & $\phantom{-}0.97$ & \\
 & \ACP & $-0.13^{+0.14}_{-0.14}$ & $\phantom{-}0.00$ & $\phantom{-}1.00$ & $\phantom{-}0.00$ \\
 & \BFratio $\times 10^2$ & $\phantom{-}4.79^{+0.72}_{-0.66}$ & $\phantom{-}0.97$ & $\phantom{-}0.00$ & $\phantom{-}1.00$ \\
\hline
\multirow{3}{0.19\textwidth}{$17.0 < q^2 < 19.0$} & \diffBF $\times 10^9$ & $\phantom{-}0.68^{+0.11}_{-0.10}$ & $\phantom{-}1.00$ & $\phantom{-}0.04$ & $\phantom{-}0.98$ & \\
 & \ACP & $\phantom{-}0.19^{+0.15}_{-0.16}$ & $\phantom{-}0.04$ & $\phantom{-}1.00$ & $\phantom{-}0.04$ \\
 & \BFratio $\times 10^2$ & $\phantom{-}3.47^{+0.59}_{-0.54}$ & $\phantom{-}0.98$ & $\phantom{-}0.04$ & $\phantom{-}1.00$ \\
\hline
\multirow{3}{0.19\textwidth}{$19.0 < q^2 < 22.0$} & \diffBF $\times 10^9$ & $\phantom{-}0.47^{+0.09}_{-0.08}$ & $\phantom{-}1.00$ & $\phantom{-}0.09$ & $\phantom{-}0.97$ & \\
 & \ACP & $\phantom{-}0.02^{+0.17}_{-0.17}$ & $\phantom{-}0.09$ & $\phantom{-}1.00$ & $\phantom{-}0.08$ \\
 & \BFratio $\times 10^2$ & $\phantom{-}5.87^{+1.10}_{-1.00}$ & $\phantom{-}0.97$ & $\phantom{-}0.08$ & $\phantom{-}1.00$ \\
\hline
\multirow{2}{0.19\textwidth}{$22.0 < q^2 < 25.0$} & \diffBF $\times 10^9$ & $\phantom{-}0.36^{+0.08}_{-0.08}$ & $\phantom{-}1.00$ & $\phantom{-}0.09$ & & \\
 & \ACP & $\phantom{-}0.19^{+0.21}_{-0.22}$ & $\phantom{-}0.09$ & $\phantom{-}1.00$ &  \\
\label{tab:bintest_stats}
\end{tabular}
\end{table}

\clearpage



\addcontentsline{toc}{section}{References}
\bibliographystyle{LHCb/LHCb}
\bibliography{main,LHCb/standard, LHCb/LHCb-PAPER, LHCb/LHCb-CONF,LHCb/LHCb-DP,LHCb/LHCb-TDR}
 
\newpage
\centerline
{\large\bf LHCb collaboration}
\begin
{flushleft}
\small
R.~Aaij$^{39}$\lhcborcid{0000-0003-0533-1952},
M.~Abdelfatah$^{71}$,
A.S.W.~Abdelmotteleb$^{59}$\lhcborcid{0000-0001-7905-0542},
C.~Abellan~Beteta$^{53}$\lhcborcid{0009-0009-0869-6798},
F.~Abudin\'en$^{61}$\lhcborcid{0000-0002-6737-3528},
T.~Ackernley$^{63}$\lhcborcid{0000-0002-5951-3498},
A.A.~Adefisoye$^{71}$\lhcborcid{0000-0003-2448-1550},
B.~Adeva$^{49}$\lhcborcid{0000-0001-9756-3712},
M.~Adinolfi$^{57}$\lhcborcid{0000-0002-1326-1264},
P.~Adlarson$^{87,44}$\lhcborcid{0000-0001-6280-3851}
% ...

{\footnotesize \it

% ...
$^{39}$Nikhef National Institute for Subatomic Physics, Amsterdam, Netherlands\\
% ...
$^{44}$National Center for Nuclear Research (NCBJ), Warsaw, Poland\\
% ...
$^{49}$Instituto Galego de F{\'\i}sica de Altas Enerx{\'\i}as (IGFAE), Universidade de Santiago de Compostela, Santiago de Compostela, Spain\\
% ...
$^{53}$Physik-Institut, Universit{\"a}t Z{\"u}rich, Z{\"u}rich, Switzerland\\
% ...
$^{57}$H.H. Wills Physics Laboratory, University of Bristol, Bristol, United Kingdom\\
% ...
$^{59}$Department of Physics, University of Warwick, Coventry, United Kingdom\\
% ...
$^{61}$School of Physics and Astronomy, University of Edinburgh, Edinburgh, United Kingdom\\
$^{62}$School of Physics and Astronomy, University of Glasgow, Glasgow, United Kingdom\\
$^{63}$Oliver Lodge Laboratory, University of Liverpool, Liverpool, United Kingdom\\
% ...
$^{71}$Syracuse University, Syracuse, NY, United States\\
% ...
$^{87}$Department of Physics and Astronomy, Uppsala University, Uppsala, Sweden, associated to $^{62}$\\
% ...
}
\end{flushleft}

\end{document}